\documentclass[11pt,a4paper]{article}

\usepackage{latexsym}

\begin{document}

\begin{center}
\begin{Large}\textbf{Nonlinear Brownian Motion and Higgs
Mechanism}\end{Large}
\begin{verbatim}
\end{verbatim}
\textit{Alexander Gl\"{u}ck \ \& \ Helmuth H\"{u}ffel}\\
Faculty of Physics, University of Vienna\\
1090 Vienna, Austria
\end{center}
\begin{abstract}
An extension of the stochastic quantization scheme is proposed by
adding nonlinear terms to the field equations. Our modification is
motivated by the recently established theory of active Brownian
motion. We discuss a way of promoting this theory to the case of
infinite degrees of freedom. Equilibrium distributions can be
calculated exactly and are interpreted as path integral densities of
quantum field theories. By applying our procedure to scalar QED, the
symmetry breaking potential of the Higgs mechanism arises as the
equilibrium solution.
\end{abstract}

\section{Introduction}
At the beginning we want to give a short review of the
interconnection between Brownian motion and Stochastic Quantization.
The dynamical evolution of the position $\textbf{q}$ of a Brownian
particle is described by the Langevin equation
(\cite{gard},\cite{arn})
\begin{equation} \label{anfang}
\frac{dq^i (t)}{dt} = - \delta^{ij} \frac{\partial U(q)}{\partial
q^j} + \eta^i (t)
\end{equation}
with
\begin{equation} \label{corr}
\langle \eta^i (t) \rangle = 0  \quad , \quad \langle\eta^i (t)
\eta^j (\bar{t}) \rangle = 2 \delta^{ij} \delta (t-\bar{t})
\end{equation}
and $U(q)$ being the external potential. The "noise"-variable $\eta$
represents the random impacts on the particle, resulting from the
molecules of the liquid, in which the particle is suspended. This
corresponds to the correlation relation (\ref{corr}), from which
follows that $\eta$ has infinite variance. The evolution of the
probability distribution $\rho (q,t)$ is given by the Smoluchowski
equation
\begin{equation}
\frac{\partial \rho (q,t)}{\partial t} = \frac{\partial}{\partial
q^i} \delta^{ij} \left(\frac{\partial U (q)}{\partial q^j} +
\frac{\partial}{\partial q^j} \right) \rho (q,t).
\end{equation}
Solving this equation for $\frac{\partial \rho }{\partial t} = 0$
defines an equilibrium distribution, which is
\begin{equation} \label{rho}
\rho_{equ} (q) \sim e^{-U(q)},
\end{equation}
so that equilibrium averages for a function $f(q)$ are calculated as
\begin{equation}\label{equ}
\langle f(q) \rangle_{equ} \sim \int d^n q \ e^{-U(q)} \ f(q).
\end{equation}
\\ 
Stochastic dynamics leading to the equilibrium solution (\ref{rho})
can be formulated also in phase space, by imposing the Langevin
equations
\begin{equation}
\frac{dq^i(t)}{dt} = \frac{\partial H (q,p)}{\partial p_i (t)} \quad
, \quad \frac{dp_i (t)}{dt} = - \frac{\partial H (q,p)}{\partial q^i
(t)} -   \delta_{ij} \frac{\partial H (q,p)}{\partial p_j (t)} +
\eta_i
\end{equation} 
with
\begin{equation}
\langle \eta_{i} (t) \rangle = 0  \quad , \quad \langle\eta_i (t)
\eta_j (\bar{t}) \rangle = 2 \delta_{ij} \delta (t-\bar{t})
\end{equation}
and $H(q,p)=\frac{p^2}{2} + U(q)$. The corresponding Fokker Planck
equation for the distribution $\rho(q,p)$ has the equilibrium solution
\begin{equation}
\rho_{equ} (q,p) \sim e^{-\frac{p^2}{2} - U(q)} = e^{-H(q,p)}.
\end{equation}
Hence, the equilibrium average for a function $f(q,p)$ is given by
\begin{equation}
\langle f(q,p) \rangle_{equ} \sim \int \ d^n q \ d^n p \  e^{-H(q,p)}
\ f(q,p).
\end{equation}
For functions $f=f(q)$, the momentum integral can be carried out by
Gaussian integration, so that
\begin{equation}
\langle f(q) \rangle_{equ} \sim \int \ d^n q \ e^{-U(q)} \ f(q)
\end{equation}
which is identical to (\ref{equ}). The idea of Stochastic
Quantization (\cite{par and wu},\cite{stoch1},\cite{stoch2}) is to
view the path integral density $e^{-S[\phi]}$ as the equilibrium
distribution of a specific stochastic process, just as $e^{-U(q)}$ is
the equilibrium distribution of the stochastic process
(\ref{anfang}). The field theoretic generalization of equation
(\ref{anfang}) is
\begin{equation} \label{lang}
\frac{\partial \phi(x,s)}{\partial s} = - \frac{\delta S}{\delta
\phi} (x,s) + \eta (x,s),
\end{equation}
where $x=(t,\textbf{x})$. $S=S[\phi]$ denotes the Euclidean action of
the theory. The noise correlations are
\begin{equation}
\langle \eta(x,s) \rangle = 0 \quad,\quad \langle  \eta (x,s) \eta
(\bar{x},\bar{s}) \rangle = 2 \delta^4 (x-\bar{x}) \delta (s-\bar{s}).
\end{equation}
Fields evolve in the parameter $s$ ('stochastic time'), which is the
evolution parameter of the stochastic process. Equation (\ref{lang})
defines a field equation for a five dimensional classical field
theory, which differs from the usual field equations  $\frac{\delta
S}{\delta \phi} =0$ by the existence of the noise field $\eta (x,s)$
and the evolution in the stochastic time.
\\ \\
The Smoluchowski equation for the probability distribution $\rho
[\phi,s]$ corresponding to the Langevin equation (\ref{lang}) is
\begin{equation} \label{fp}
\frac{\partial \rho [\phi,s]}{\partial s} = \int d^4 x \
\frac{\delta}{\delta \phi} \left(  \frac{\delta S[\phi]}{\delta \phi}
+ \frac{\delta}{\delta \phi},    \right) \rho [\phi,s]
\end{equation}
whose equilibrium solution can be read of directly:
\begin{equation}
\rho_{equ} [\phi] \sim e^{-S[\phi]}.
\end{equation}
Quantum field theory Green functions can thus be interpreted as the
equilibrium limit of stochastic averages:
\begin{equation}
G(x_1, \dots ,x_n) = \int \mathcal{D} \phi \ \rho_{equ} [\phi] \ \phi
(x_1) \dots \phi (x_n).
\end{equation}

\section{Active Brownian Motion in Phase Space}

A possible extension of the Brownian motion model is given by
introducing nonlinear terms in the dynamical equations. Nonlinear
Brownian Motion is particulary described by the theory of so-called
\textit{active} Brownian motion, which was originally formulated in
phase space (\cite{schw und ebe},\cite{erdm},\cite{schw},\cite{ebe
und sokol}). Active Brownian particles differ from ordinary Brownian
particles by an additional internal
degree of freedom $e$, which is interpreted as a form of internal
energy. The $2n$-dimensional phase space is extended by introducing a
new internal degree of freedom $e$, which satisfies the following
dynamical equation:
\begin{equation}
\frac{de}{dt}=c_{1}-c_{2}e-c_{3}e\frac{p^2}{2}
\end{equation}
The parameters $c_{1}$,$c_{2}$,$c_{3}$ are assumed to be constant and
positive. A physical motivation for the equation above can be given
in terms of biological systems. $c_{1}$ represents a constant intake
of energy, the second term describes a loss of energy due to internal
processes, and the last term models the coupling of $e$ to the
momentum, in the form of a conversion of internal energy into kinetic
energy. Active Brownian particles can thus be viewed as organisms
with an internal energy depot, which is increased by taking up food
(first term), and decreased by internal processes of the organism
(second term) and conversion of internal energy into motion (third
term).
\\ \\
Stochastic dynamics in the extended phase space are closed by
specifying a coupling of the coordinates $(\textbf{q},\textbf{p})$ to
the internal energy:
\begin{equation}
\frac{dq^i}{dt}=\frac{\partial H}{\partial p_{i}}
\end{equation}
\begin{equation}
\frac{dp_{i}}{dt}=-\frac{\partial H}{\partial q^i}-
\delta_{ij}\frac{\partial H}{\partial p_j}+c e
\delta_{ij}\frac{\partial H}{\partial p_j}+\eta_i
\end{equation}
with
\begin{equation}
\langle\eta_i (t) \rangle = 0   \quad , \quad \langle\eta_i (t)
\eta_j (\bar{t}) \rangle = 2 \delta_{ij} \delta (t-\bar{t})
\end{equation}
where $H=\frac{p^2}{2}+U(q)$ and $c>0$. Assuming that the evolution
of the internal energy takes place on an essentially shorter time
scale than that of the other coordinates, we can set
$\frac{de}{dt}=0$ and thus get an explicit depence of the internal
energy on the momentum:
\begin{equation}
e=e(p)=\frac{c_1}{c_2 + c_3 \frac{p^2}{2}}.
\end{equation}
Hence, the internal energy can be eliminated from the dynamical
equations and we are left with stochastic dynamics that represent a
system with a momentum dependent friction function:
\begin{equation}
\frac{dq^i}{dt}=\frac{\partial H}{\partial p_{i}}
\end{equation}
\begin{equation}
\frac{dp_{i}}{dt}=-\delta_{ij}\frac{\partial H}{\partial
q^i}-\gamma(p) \delta_{ij}\frac{\partial H}{\partial p_j}+\eta_i
\end{equation}
\begin{equation}
\gamma (p) = 1- c \frac{c_1}{c_2 + c_3 \frac{p^2}{2}}.
\end{equation}
Given the assumption that the coupling of $e$ to the kinetic energy
is small ($c_3\ll1$), an expansion to first order in $p^2$ yields
$\gamma (p) \simeq \gamma_{1}+\gamma_{2} p^2$ with \linebreak
$\gamma_{1} \equiv \gamma_{0} - c\frac{c_1}{c_2}$ and $\gamma_{2}
\equiv c \frac{c_1 c_3}{2 {c_2}^2}$. $\gamma_1$ and $\gamma_2$ can be
chosen independently of each other. $\gamma_2$ is strictly positive,
whereas $\gamma_1$ can be chosen to be negative. Explicit
computations of equilibrium solutions in phase space are done by
specifying the potential $U(q)$ and by carrying out a certain avering
procedure, which shows up to be necessary due to the nonlinear nature
of the equations. Given the case of a harmonic potential and one
degree of freedom: $H=\frac{p^2}{2}+w^2\frac{q^2}{2}$, the stochastic
differential equations are
\begin{equation}
\frac{dq}{dt}=p \quad,\quad \frac{dp}{dt}=-w^2 q
-(\gamma_{1}+\gamma_{2} p^2)p + \eta.
\end{equation}
Transforming via the Ito formula to action-angle variables $(H,\phi)$:
\begin{equation}
q(H,\phi)=\frac{\sqrt{2H}}{w} \cos{\phi} \quad,\quad
p(H,\phi)=\sqrt{2H} \sin{\phi}
\end{equation}
and taking averages with respect to $\phi$, an averaged Fokker Planck
equation for the distribution $\rho[H,t]$ can be obtained: 
\begin{equation}
\frac{d\rho}{dt}=\frac{\partial}{\partial H}\left(\gamma_{1} H +
\frac{3}{2}\gamma_{2} H^2 +H\frac{\partial}{\partial H} \right)\rho,
\end{equation}
whose equilibrium solution reads
\begin{equation} \label{one dim solution}
\rho_{equ} (H) \sim e^{aH-bH^2}
\end{equation} 
with $a\equiv - \gamma_1 > 0$ and $b\equiv \frac{3}{4} \gamma_2 >0$
being the independent parameters of the theory. It is remarkable that
there is no direct way of generalizing this result to higher degrees
of freedom. In \cite{mao}, the result for $n=2$ was computed to be
\begin{equation}
\rho_{equ} (H_1,H_2,\theta) \sim e^{a(H_1 + H_2) - b \lbrace
{(H_1)}^2+{(H_2)}^2 \rbrace+ \frac{4}{3} b H_1 H_2 \lbrace
1+\frac{1}{2}  \cos (2\theta)   \rbrace},
\end{equation}
where $H_i = \frac{{p_i}^2}{2} + w^2 \frac{{q^i}^2}{2}$; $a,b>0$ and
$\theta$ being an angle variable coming from the avering procedure.

\section{Active Brownian Motion in Configuration Space}

We now give a new formulation of active Brownian motion in
$\textbf{q}$-space and calculate an \textit{exact} equilibrium
distribution.Whereas in the phase space formalism a coupling of the
internal energy to the momentum was proposed, we now specify a
coupling to the position $\textbf{q}$. A natural choice is to impose
a coupling of $e$ to the potential energy $U(q)$, in analogy to the
formalism in phase space, where a coupling of the internal energy to
the kinetic energy was stated. We will work with a general potential
$U(q)$ in what follows, which is a main difference to the
calculations in phase space, where the potential has to be specified
in order to calculate an approximative equilibirum distribution. In
analogy to the situation in phase space, where $e$ was coupled to the
momentum-gradient of the mechanical energy, we now postulate a
coupling of $e$ to the $\textbf{q}$-gradient of
$H(\textbf{q},\textbf{p})$. Stochastic dymanics are given by the
following equations:
\begin{equation} \label{coupling in config space}
\frac{dq^{i}}{dt}=-\delta^{ij}\frac{\partial U}{\partial q^j}+c e
\delta^{ij}\frac{\partial U}{\partial q^j} +\eta^i
\end{equation}
\begin{equation} \label{energy}
\frac{de}{dt}=c_{1}-c_{2}e-c_{3}e U(q)
\end{equation}
with
\begin{equation} \label{noise}
\langle \eta^i (t) \rangle = 0 \quad , \quad \langle\eta^i (t) \eta^j
(\bar{t}) \rangle = 2 \delta^{ij} \delta (t-\bar{t}).
\end{equation}
$c,c_1,c_2,c_3$ are assumed to be positive constants. Elimination of
$e$ allows us to calculate an exact equilibrium solution of the
Fokker Planck equation. We set $\frac{de}{dt}=0$, so that $e$ is
expressed in terms of the absolute value of the position:
\begin{equation}
e=e(q)=\frac{c_1}{c_2 + c_3 U(q)}.
\end{equation}
Plugging this into (\ref{coupling in config space}) shows that the
coupling term can be written as a gradient. Therefore, the Langevin
equation is in the standard form
\begin{equation}
\frac{dq^{i}}{dt}=-\delta^{ij}\frac{\partial \tilde{U}}{\partial
q^j}+\eta^i
\end{equation}
with the effective potential
\begin{equation} \label{modified potential}
\tilde{U}(q)=U(q)-c\frac{c_1}{c_3} \ln \left\lbrace c_2+c_3 U(q)
\right\rbrace
\end{equation}
The Smoluchowski equation reads
\begin{equation}
\frac{d\rho}{dt}=\frac{\partial}{\partial q^i} \delta^{ij}
\left(\frac{\partial  \tilde{U}   }{\partial q^j}  +
\frac{\partial}{\partial q^j}  \right)\rho,
\end{equation}
which has the equilibrium solution
\begin{equation} \label{q-space solution}
\rho_{equ} (q) \sim e^{- \tilde{U} (q)} = e^{-U(q)+c\frac{c_1}{c_3}
\ln \left\lbrace c_2 + c_3 U(q)  \right\rbrace}.
\end{equation}
\begin{flushleft}
There are three remarks:
\end{flushleft}
\begin{itemize}
\item No averaging procedure was needed to find an equilibrium
distribution, like in the previous phase space formalism.
\item The result is valid for a general dimension $n$. We have
mentioned the fact that in the phase space formalism, generalizations
of the one dimensional case to higher dimensions are nontrivial. In
the present formulation, we do not have to restrict ourselves to
lower dimensional cases.
\item The above computations hold for a \emph{general} potential
$U(q)$, whereas in the action-angle based averaging procedure in
phase space the potential has to be specified explicitly.
\end{itemize}
Factorizing the logarithm in (\ref{q-space solution}) and absorbing
the factor $\ln(c_2)$ into the normalization, we expand the remaining
term under the assumption that the coupling of $e$ to the potential
energy is small compared to internal dissipation ($\frac{c_3}{c_2}$
small). By choosing $1-c\frac{c_1}{c_3}<0$ and with the definitions
$a\equiv -(1-c\frac{c_1}{c_3})$ and $b \equiv \frac{1}{2} c \frac{c_1
c_3}{(c_2)^2}$, the effective potential reads
\begin{equation}
\tilde{U}(q) = aU(q)-bU(q)^2
\end{equation}
with $a,b>0$ being the independet parameters of the theory. Given the
case of one dimension and by assuming a harmonic potential $U(q)=w^2
\frac{q^2}{2}$, $\tilde{U} (q)$ has two minima at ${q_0}_{\pm} = \pm
\frac{1}{w} \left( \frac{a}{b} \right)^{\frac{1}{2}}$. Studying this
theory near one of the minima yields the classical mechanism of
symmetry breaking, which means that active Brownian motion in
configuration space under the assumption of a harmonic external
potential gives rise to an equilibrium distribution with broken
symmetry.

\section{Nonlinear Stochastic Quantization}

We set up a field theory generalization of active Brownian motion in
configuration space and calculate an exact equilibrium distribution.
When applied to scalar QED, this modified stochastic quantization
procedure leads to the symmetry breaking potential of the Higgs
mechanism. \\ \\
Equations (\ref{coupling in config space})-(\ref{noise}) for active
Brownian motion in configuration space are generalized to
\begin{equation} \label{config space fields}
\frac{\partial \phi^i (x,s)}{\partial s} = - \delta^{ij}
\frac{\delta S}{\delta \phi^j }(x,s) + ce \delta^{ij} \frac{\partial
V}{\partial \phi^j }(x,s) + \eta^i (x,s)
\end{equation}
\begin{equation}
\frac{\partial e}{\partial s} = c_1 - c_2 e - c_3 e V(\phi)
\end{equation}
\begin{equation}
\langle \eta(x,s)^i \rangle = 0 \quad,\quad \langle  \eta(x,s)^i
\eta(\bar{x},\bar{s})^j \rangle = 2\delta^{ij} \delta^4
(x-\bar{x})\delta (s-\bar{s} )
\end{equation}
for a collection of fields $\lbrace \phi^i \rbrace = \phi$, where
$S=S[\phi]$ denotes the Euclidean action and the constants
$c,c_1,c_2,c_3$ are assumed to be positive. Setting $\frac{\partial
e}{\partial s} = 0$, we get an expression of $e$ in terms of the
fields:
\begin{equation}
e=e(\phi)=\frac{c_1}{c_2 + c_3 V(\phi)}.
\end{equation}
Plugging this into (\ref{config space fields}), the field equations
can be written in the standard form
\begin{equation} \label{standard form}
\frac{\partial \phi^i (x,s)}{\partial s} = - \delta^{ij}
\frac{\delta \tilde{S}}{\delta \phi^j } (x,s) + \eta^i (x,s)
\end{equation}
with the effective action
\begin{equation}
\tilde{S} [\phi] \equiv \int d^4 x \left\lbrace
\mathcal{L}(\phi,\partial\phi) - \frac{c c_1}{c_3} \ln \Big(c_2 + c_3
V(\phi) \Big) \right\rbrace = \int d^4 x \
\tilde{\mathcal{L}}(\phi,\partial\phi).
\end{equation}
The effective Lagrangian $\tilde{\mathcal{L}}$ is analogous to the
effective potential (\ref{modified potential}). The Smoluchowski
equation corresponding to (\ref{standard form}) reads
\begin{equation}
\frac{\partial \rho[\phi,s]}{\partial s} = \int d^4 x
\frac{\delta}{\delta \phi^i} \delta^{ij} \left( \frac{\delta
\tilde{S} }{\delta \phi ^j} + \frac{\delta}{\delta \phi^j} \right)
\rho [\phi,s],
\end{equation}
which has the equilibrium solution
\begin{equation}
\rho_{equ} [\phi] \sim e^{-\tilde{S} [\phi]}.
\end{equation}
Carrying out the expansion of the effective Lagrangian in the same
way as we did in the case of finite degrees of freedom leads to an
effective field theory potential of the form
\begin{equation}
\tilde{V}(\phi)=-aV(\phi)+bV(\phi)^2
\end{equation}
with $a=-(1-c \frac{c_1}{c_2}) > 0$ and $b = \frac{1}{2} c \frac{c_1
c_3}{(c_2)^2} > 0$. Equilibrium Green functions are then given by
\begin{equation}
G(x_1,\dots,x_n) \sim \int \mathcal{D} \phi \ e^{-\int d^4 x\lbrace
\frac{1}{2} ({\partial \phi})^2- aV(\phi) + b V(\phi)^2    \rbrace }
\  \phi(x_1) \dots \phi(x_n)
\end{equation}
and are identified with the Green functions of a quantized field
theory. Applying the general formalism outlined above in the case of
a free,  real scalar field demonstrates how symmetry breaking
immediately drops out of the nonlinear stochastic quantization
scheme. The Euclidean Lagrangian
\begin{equation}
\mathcal{L} (\phi,\partial\phi) = \frac{1}{2} (\partial \phi)^2 +
\frac{m^2}{2} \phi^2
\end{equation}
leads to the equilibrium distribution
\begin{equation} \label{equ distr scalar}
\rho_{equ} [\phi] \sim e^{-   \int d^4 x  \left\lbrace \frac{1}{2}
(\partial\phi)^2 - a\frac{m^2}{2}\phi^2 + b \frac{m^4}{4} \phi^4
\right\rbrace   }
\end{equation}
with $a,b>0$. Identifying the effective potential
\begin{equation}
\tilde{V}(\phi)=-\frac{a}{2} m^2 \phi^2 + \frac{b}{4} m^4 \phi^4
\end{equation}
and its minima
\begin{equation}
\phi_{0_{\pm}} = \pm \frac{1}{m} \left( {\frac{a}{b}} \right)
^{\frac{1}{2}},
\end{equation}
the theory can be studied near one of its minima in the usual way.
Having this simple example in mind, we can proceed to the
construction of the Higgs mechanism via the application of the
nonlinear stochastic quantization scheme to scalar QED. The Euclidean
Lagrangian
\begin{equation}
\mathcal{L}=\frac{1}{4}F_{\mu\nu}F^{\mu\nu}+
(D_{\mu}\phi)^{\ast}(D^{\mu}\phi)+ V(\vert\phi\vert) \quad, \quad
V(\vert\phi\vert)  =  m^2\phi^{\ast}\phi
\end{equation}
($D=\partial - iA$) leads to the following stochastic differential
equations:
{\setlength\arraycolsep{2pt}
\begin{eqnarray}
\frac{\partial A_{\mu}}{\partial s} & = & \partial^{\nu} F_{\nu\mu} -
i(\phi^{\ast} \partial_{\mu} \phi - \phi \partial_{\mu} \phi^{\ast}
) - 2 \phi^{\ast} \phi A_{\mu}  + %{} \nonumber  \\ && {}  
 \xi_{\mu}
\\
\frac{\partial\phi}{\partial s} & = &  \lbrace (D_{\mu}
D^{\mu})^{\ast} - m^2  \rbrace \phi^{\ast} +   ce m^2\phi^{\ast}+ \eta
\\
\frac{\partial\phi^{\ast}}{\partial s} & = & \lbrace D_{\mu} D^{\mu}
-m^2 \rbrace \phi +  ce m^2\phi + \eta^{\ast}
\\
\frac{\partial e}{\partial s} & = & c_1 - c_2 e - c_3 e m^2 \phi
^{\ast} \phi 
\end{eqnarray}}
with the noise correlations
\begin{equation}
\langle \eta(x,s) \rangle = 0\quad ,\quad \langle \eta(x,s)^{\ast}
\rangle = 0 \quad,\quad \langle  \xi_{\mu} (x,s) \rangle = 0
\end{equation}
\begin{equation}
\langle \eta (x,s) \eta (\bar{x},\bar{s}) ^{\ast}  \rangle = 2 \delta
^4 (x-\bar{x}) \delta (s-\bar{s}) 
\end{equation}
\begin{equation}
\langle  \xi_{\mu}(x,s)  \xi_{\nu} (\bar{x},\bar{s}) \rangle = 2
\delta_{\mu\nu} \delta^4 (x-\bar{x})\delta(s-\bar{s})
\end{equation}
Elimination of the internal energy finally allows us to write this
set of equations in the form
\begin{equation}
\frac{\partial\varphi^i(x,s)}{\partial
s}=-\delta^{ij}\frac{\delta\tilde{S}}{\delta\varphi ^{j}} (x,s) +
\chi^{i} (x,s)
\end{equation}
with
\begin{equation}
\langle \chi^i (x,s)  \rangle = 0 \quad,\quad \langle \chi^i (x,s)
\chi^j (\bar{x},\bar{s} )  \rangle = 2 \delta ^{ij} \delta^4
(x-\bar{x}) \delta (s-\bar{s})
\end{equation}
where $\varphi=(\phi,\phi^{\ast},A)$ and $\chi =
(\eta,\eta^{\ast},\xi)$. Approximating the effective action like in
the previous examples yields
\begin{equation}
\tilde{S}[\varphi]=\int d^4 x \left \lbrace   \frac{1}{4}
F_{\mu\nu}F^{\mu\nu}+ (D_{\mu}\phi)^{\ast}(D^{\mu}\phi) - am^2 \vert
\phi \vert^2 +bm^4 \vert \phi \vert ^4   \right \rbrace.
\end{equation}
In complete analogy to the stochastic quantization procedure for QED
(\cite{huffel}) a gauge fixed equilibrium distribution can be
obtained straightforwardly:
\begin{equation}
\rho_{equ} \sim  e^{-\int d^4 x \ \tilde{\mathcal{L}}_{gf} },
\end{equation}
where the gauge fixed effective Lagrangian reads
\begin{equation}
\tilde{\mathcal{L}}_{gf} =  \frac{1}{4} F_{\mu\nu}F^{\mu\nu}+
\frac{1}{2} (\partial A)^2  + (D_{\mu}\phi)^{\ast}(D^{\mu}\phi) -
am^2 \vert \phi \vert^2 +bm^4 \vert \phi \vert ^4 .
\end{equation}
The effective potential
\begin{equation}
\tilde{V}(\vert\phi\vert)=-am^2 \vert\phi\vert^2+ b m^4
\vert\phi\vert ^4
\end{equation}
has two minima at
\begin{equation}
\vert\phi\vert_{0_{\pm}} = \pm \frac{1}{m} \left( {\frac{a}{2b}}
\right) ^{\frac{1}{2}}
\end{equation}
Shifting the Lagrangian, the mass of the gauge field and the Higgs
mass are identified:
\begin{equation}
{m_A}^2 = \frac{a}{b}\left(\frac{1}{m}\right)^2 \quad,\quad
{m_{H}}^2 = 2am^2
\end{equation}
\\
\section{Conclusion}
We would like to emphasize the following features of our procedure
that distinguishes it from the usual way of constructing the Higgs
mechanism:
\begin{itemize}
\item Performing the nonlinear stochastic quantization of scalar QED
in the approximation of a small coupling of the potential to an
internal energy, we arrive at an effective potential equivalent to
the symmetry breaking potential usually introduced in the theory of
Higgs mechanism. Our starting point was ordinary scalar QED with a
potential $V(\vert\phi\vert)\sim \vert\phi\vert^2$ and there was no
need to introduce a quartic interaction by hand, as is usually done.
\item The parameters $a,b$ of the effective potential have origins in
the dynamical structure of the internal energy and its coupling to
the potential $V(\vert\phi\vert)$.
\end{itemize}

\end{document}